\documentclass[prb,twocolumn,showpacs]{revtex4}

\usepackage{graphicx}
\usepackage{dcolumn}
\usepackage{amsmath}
\usepackage{color}

\newcommand{\YBa}{YBaMn$_2$O$_6$}

\newcommand{\LaSr}{La$_{1-x}$Sr$_x$MnO$_3$}

\newcommand{\etal}{\textit{et al.~}}

\begin{document}
\title{Electron Spin Resonance across the Charge-ordering Transition in YBaMn$_2$O$_6$}

\author{D.~Zakharov}
\affiliation{Experimentalphysik V, Center for Electronic
Correlations and Magnetism, Institute for Physics, Augsburg
University, D-86135 Augsburg, Germany}

\author{J.~Deisenhofer}
\affiliation{Experimentalphysik V, Center for Electronic
Correlations and Magnetism, Institute for Physics, Augsburg
University, D-86135 Augsburg, Germany}

\author{H.-A.~Krug von Nidda}
\affiliation{Experimentalphysik V, Center for Electronic
Correlations and Magnetism, Institute for Physics, Augsburg
University, D-86135 Augsburg, Germany}

\author{T.~Nakajima}
\affiliation{Material Design and Characterization Laboratory,
Institute for Solid State Physics, University of Tokyo, 5-1-5
Kashiwanoha, Kashiwa, Chiba 277-8581, Japan}

\author{Y.~Ueda}
\affiliation{Material Design and Characterization Laboratory,
Institute for Solid State Physics, University of Tokyo, 5-1-5
Kashiwanoha, Kashiwa, Chiba 277-8581, Japan}

\author{A.~Loidl}
\affiliation{Experimentalphysik V, Center for Electronic
Correlations and Magnetism, Institute for Physics, Augsburg
University, D-86135 Augsburg, Germany}

\date{\today}

\begin{abstract}
We investigated the metal-ordered manganite system YBaMn$_2$O$_6$
using electron spin resonance (ESR) in the paramagnetic regime
across the charge-ordering and structural phase transition at
$T_{\rm CO}$=480~K and $T_{\rm t}$=520~K, respectively. All ESR
parameters exhibit jump-like changes at $T_{\rm t}$ while the
charge-ordering at $T_{\rm CO}$ manifests itself only as a weak and
broad anomaly. Above $T_{\rm t}$ the ESR spin susceptibility is
reduced with respect to the $dc$-susceptibility, indicating that
only the $t_{\rm 2g}$-core spins of Mn ions contribute to the
resonance absorption. The contribution of the $e_{\rm g}$-spins is
suppressed by the time scale of the polaronic hopping process of the
$e_{\rm g}$-electrons. The linewidth in this regime is reminiscent
of a Korringa-type relaxation behavior. In this picture the ESR
properties below $T_{\rm t}$ are dominated by the slowing down of
the polaronic hopping process. The charge fluctuations persist down
to the temperature $T^* \approx 410$~K, below which the system can
be described as a charge-ordered assembly of Mn$^{3+}$ and Mn$^{4+}$
spins.
\end{abstract}


\pacs{76.30.-v, 71.70.Ej, 75.30.Et, 75.30.Vn}

\maketitle

\section{Introduction}

\begin{figure}[b]
\centering
\includegraphics[width=70mm,clip]{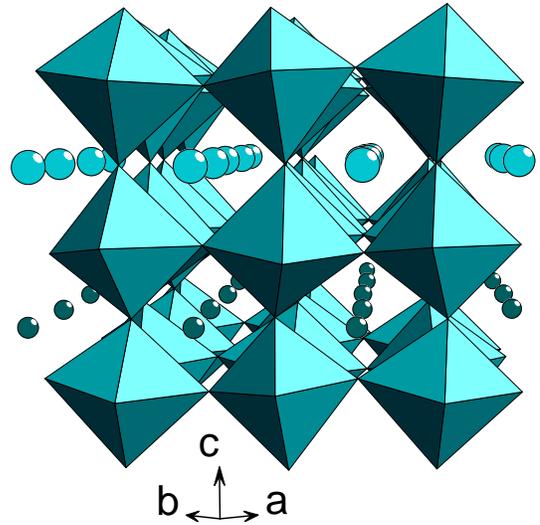}
\vspace{2mm} \caption[]{\label{Struclay} Crystal structure of \YBa.
The Y ions (small spheres) and the Ba ions (large spheres) are
arranged in alternating layers separated by MnO$_6$ octahedra.}
\end{figure}

Quenched disorder has recently been singled out to be one of the
necessary ingredients to understand the complex phase diagrams of
manganite systems and the appearance of phenomena like colossal
magnetoresistance
(CMR).\cite{Akahoshi03,Nakajima04,Dagotto01,Deisenhofer05} In most
of these compounds $R^{3+}_{x}A^{2+}_{1-x}$MnO$_3$ with $R$ being a
rare-earth element and $A$ being an alkaline-earth element such
disorder inevitably exists due to the random distribution of
$R^{3+}$ and $A^{2+}$ ions in the lattice. In order to sort out
which of the plethora of phenomena in the phase diagrams are due to
disorder and which are not, a controlled occupation of the
corresponding lattice sites has to be achieved. This task has been
accomplished in the case of half substituted systems
$R^{3+}_{0.5}A^{2+}_{0.5}$MnO$_3$, which could be successfully
synthesized as $R$BaMn$_2$O$_6$, a new class of metal-ordered
manganites.\cite{Nakajima02b,Williams05} In contrast to the
disordered systems, in these compounds the MnO$_2$ planes are
sandwiched by two types of rock-salt layers, $R$O and BaO, with
different lattice sizes, as can be seen in Fig.~\ref{Struclay}.

The ordered compounds with $R$=(Y,Tb,...,Sm) exhibit a transition
from a metallic to a charge-ordered (CO) state within the
paramagnetic (PM) regime at temperatures as high as $T_{\rm CO} =
480$~K for $R=$ Y and an antiferromagnetic ground state. For ions
with larger ionic radii $R$=(Nd,Pr,La) ferromagnetic ordering
appears above room temperature. Indeed, it was shown that the
absence of A-site disorder drastically stabilizes the CO phase and
does prevent from the occurrence of magnetoresistance
effects.\cite{Nakajima05} In reverse, a CMR effect of about 1000\%
at room temperature was found by a subtle tailoring of the CO
transition via the ionic radius and introducing a sophisticated kind
of disorder.

In this study we focus on \YBa\; which exhibits a transition from a
charge-ordered insulating to a metallic regime at higher
temperatures within the PM regime. We investigate the spin dynamics
by electron spin resonance (ESR) spectroscopy, which has been shown
to be a very efficient tool to probe orbital order and charge order
in various transition-metal oxides, because ESR allows to directly
access the spin of interest and probe its local symmetry and
relaxation
behavior.\cite{Kochelaev03,Ivanshin03,Zakharov08,Deisenhofer03,Zakharov03}

\section{Experimental Results}

ESR measurements were performed in a Bruker ELEXSYS E500
CW-spectrometer at X-band frequencies ($\nu \approx$ 9.47 GHz)
equipped with a continuous N$_2$-gas-flow cryostat in the
temperature region $80<T< 600$ K. The polycrystalline samples were
powdered and placed into quartz tubes. ESR detects the power $P$
absorbed by the sample from the transverse magnetic microwave field
as a function of the static magnetic field $H$. The signal-to-noise
ratio of the spectra is improved by recording the derivative $dP/dH$
using lock-in technique with field modulation.

\begin{figure}[tb]
\centering
\includegraphics[width=65mm,clip,angle=0]{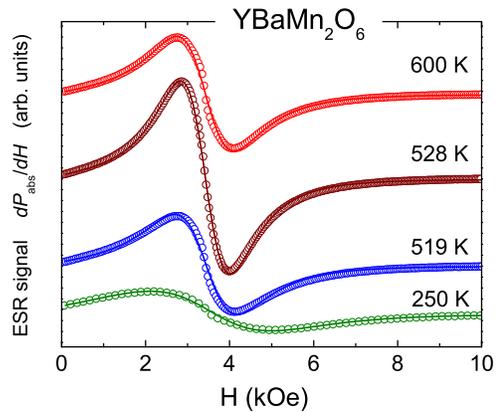}
\vspace{2mm} \caption[]{\label{Spectra}(color online) Temperature
evolution of the ESR spectrum in \YBa. Solid lines represent fits
using a Lorentzian line shape.}
\end{figure}

In Figure \ref{Spectra} we show ESR spectra in YBaMn$_2$O$_6$ at
different temperatures. The spectra consist of a broad,
exchange-narrowed resonance line, which can be very well fitted by a
single Lorentzian line shape. In order to check the origin of the
ESR signal we determined the absolute value of the spin
susceptibility by comparison with the reference compound
Gd$_2$BaCuO$_5$. It turns out, that in the temperature range below
about $T^* \approx 410$~K the susceptibility determined from ESR
$\chi_{\rm ESR}$ coincides with the $dc$-susceptibility $\chi_{dc}$
(see Fig.~\ref{YBMO}). Above $T^*$ $\chi_{\rm ESR}$ gradually
deviates from $\chi_{dc}$. The difference reaches its maximal value
at the structural transition $T_{\rm t} = 520$~K and remains
approximately constant at higher temperatures. The effective moment
determined from the $dc$-susceptibility in the high-temperature
phase $T>T_{\rm t}$ agrees nicely with $\mu_{dc}$=6.245~$\mu_B$,
\cite{Nakajima04} obtained when assuming that all
Mn$^{3+}$/Mn$^{4+}$ ions contribute to the effective moment.
However, the magnetic moment determined from the ESR measurements
above $T_{\rm t}$ is considerably smaller and amounts to $\mu_{\rm
ESR}$=5.0(3)~$\mu_B$.

In Fig.~\ref{YBMO}(a) we compare the temperature dependence of the
spin susceptibility $\chi_{\rm ESR}$ obtained by ESR and the
$dc$-susceptibility $\chi_{\rm dc}$. In general, the differences
upon heating and cooling and the features at the phase transitions
are similar for $\chi_{\rm ESR}$ and $\chi_{\rm dc}$. However, one
can clearly see the differences in the observed effective magnetic
moments in the high-temperature Curie-Weiss (CW) regime.
Interestingly, this discrepancy persists throughout the two
successive phase transitions at $T_{\rm CO} = 480$~K and $T_{\rm t}
= 520$~K. The former is visible in the susceptibility as a weak and
broad anomaly and has been dubbed the charge ordering transition as
the resistivity drops by more than one order of magnitude at 480~K.
\cite{Nakajima04} The latter is a structural transition from a
triclinic structure below 520~K to a monoclinic one
above.\cite{Williams05b} This structural transition shows up most
dramatically in the magnetic susceptibilities as a strong increase
and a subsequent renormalized CW law for $T>T_{\rm t}$, while the
changes in the resistivity at $T_{\rm t}$ are subtle.
\cite{Nakajima04}

Let us now discuss the temperature dependence of the effective
$g$-factor (Fig.~\ref{YBMO}(b)). In the high-temperature metallic
phase $g=1.98$ is slightly below the free-electron value in
accordance with the expectation for transition-metal ions with a
less than half-filled $d$ shell.\cite{Abragam70} This value
coincides with the isotropic $g$ value observed in the orbitally
disordered phase in lightly doped \LaSr\; and in metallic manganite
compounds.\cite{Causa98,Ivanshin00} Being constant above $T_{\rm
t}$, the $g$-factor then drops at the structural transition and
decreases for $T < T_{\rm t}$ down to $g=1.92$ at 200~K. The
transition at $T_{\rm CO}$=480~K appears also here only as a broad
anomaly in the temperature dependence. Below 200~K the $g$-factor
shows a steep increase which is due to internal fields in the
vicinity of magnetic ordering at $T_{\rm N}=195$~K. To check the
reliability of the obtained $g$-values, we performed ESR
measurements in Q-Band ($\nu \approx 34$~GHz, $H_{\rm res} \approx
12$~kOe) at several temperatures below 300~K which agreed well with
the X-Band measurements.

Similarly to the $g$-factor, the linewidth in \YBa~shows distinct
differences between the high-temperature metallic phase and the
charge-ordered state (Fig.~\ref{YBMO}(c)). Just above $T_{\rm t}$
the linewidth exhibits its minimum value of $\Delta H=1$~kOe and
increases linearly towards higher temperatures. The linewidth
increases sharply at $T_{\rm t}$  and rises continuously with
decreasing temperature up to 3~kOe. Again, at $T_{\rm CO}$ the
resistivity drop gives only rise to a weak and broad anomaly. A kink
is observed at the antiferromagnetic ordering temperature $T_{\rm N}
= 195$~K.

\section{Discussion}
As described above all ESR parameters and the $dc$-susceptibility
show sharp changes at $T_{\rm t}$, while $T_{\rm CO}$ leaves its
trace only as a weak and broad anomaly. It has been pointed out by
Williams \etal that in the temperature regime $T_{\rm CO} < T <
T_{\rm t}$ the system exhibits a coexistence of two
phases.\cite{Williams05b} We believe that with the appearance of
clusters of the high-temperature phase the global CO may already be
broken at $T_{\rm CO}$, conducting pathes show up which lead to the
strong drop in resistivity, but the magnetic properties are still
determined by the CO regions and vary continuously with the melting
of the CO phase towards higher temperatures. Only with the complete
disappearance of local charge-ordered areas at the first-order
structural transition at $T_{\rm t}$, the magnetic properties
undergo the most drastic changes as the exchange couplings and the
local coupling between $t_{2g}$ and $e_g$ spins become renormalized.
Therefore, we will discuss our results for $T>T_{\rm t}$ and
$T<T_{\rm t}$ separately in the following.

\subsection{High-temperature metallic phase ($T>T_{\rm t}$)}

The effective magnetic moment of about $\mu_{\rm
ESR}$=5.0(3)~$\mu_{\rm B}$ observed by ESR in the high-temperature
metallic phase is reduced with respect to the $dc$-susceptibility,
which is in agreement with the value of 6.245~$\mu_{\rm B}$ expected
for all Mn$^{3+}$ and Mn$^{4+}$ spins. Such a behavior is
astonishing, especially because in prototypical manganite systems
like La$_{1-x}$Sr$_x$MnO$_3$ the observed ESR spin susceptibilities
yielded effective magnetic moments, which were enhanced in
comparison to the sum of all Mn$^{3+}$/Mn$^{4+}$ spins.
\cite{Causa98,Ivanshin00} This effect was interpreted by Shengelaya
{\it et al.} due to a bottlenecked spin relaxation of the
exchange-coupled  Mn$^{3+}$/Mn$^{4+}$ spin system to the
lattice.\cite{Shengelaya96,Shengelaya00}

To account for the origin of our reduced ESR susceptibility one may
imagine a situation where either the Mn$^{3+}$-subsystem or the
Mn$^{4+}$ ions participate in the resonance process. In both cases,
however, an even lower effective moment is expected and,
additionally, it is difficult to reasonably justify such a scenario.
We think that in \YBa~the magnetic moment observed by ESR has to be
compared to the magnetic moment of 5.48~$\mu_B$ expected if only the
$t_{2g}$-core spins of all Mn ions are seen by ESR, i.e.~that all Mn
ions are in a tetravalent state. If this is the case, then we have
to cope with fact that we are at odds with the textbook knowledge of
a strong local Hund's coupling between the $t_{2g}$-core spins and
the $e_g$-electrons. Recently, Huber \etal~challenged this textbook
picture and discussed the possibility that the $dc$-susceptibility
in paramagnetic manganites can not only be understood as due to
Mn$^{3+}$/Mn$^{4+}$ spins, but it can equivalently be modeled by a
process where the Hund's coupling between the $t_{2g}$-core spins
($S=3/2$-system) and the spins residing in the $e_g$-state
($S=1/2$-system) is disrupted by the hopping of the $e_g$-electrons.
As a result of the thermally activated hopping the Curie constant of
the system will become temperature dependent.\cite{Huber07} Note,
that this picture can describe quantitatively the static
$dc$-susceptibility of manganite compounds including YBaMn$_2$O$_6$.

\begin{figure}[t]
\centering
\includegraphics[width=80mm,clip]{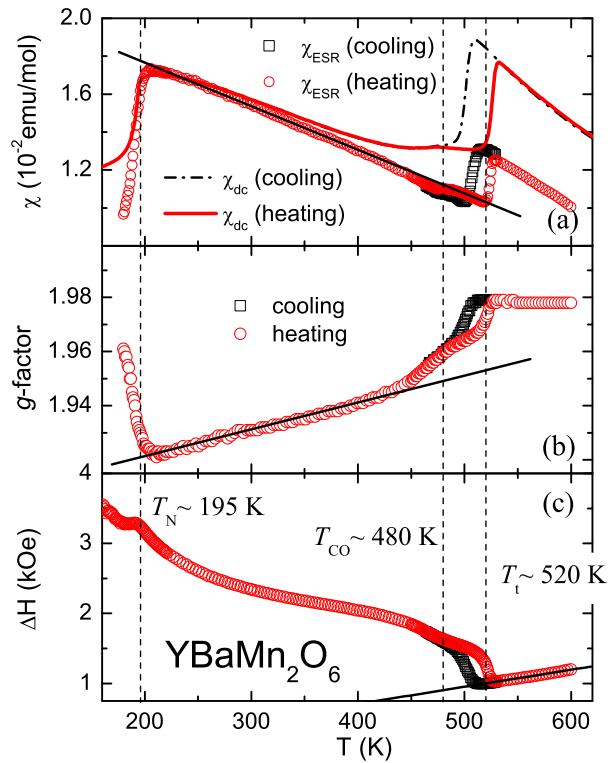}
\vspace{2mm} \caption[]{\label{YBMO}Temperature dependence of (a)
the ESR spin susceptibility $\chi_{\rm ESR}$ and the
$dc$-susceptibility $\chi_{\rm dc}$, (b) the effective $g$-factor,
and (c) the ESR linewidth $\Delta H$ in YBaMn$_2$O$_6$. The lines in
the frames (a) and (b) are to guide the eyes, the line in frame (c)
is a fit according to the Korringa relation given in the text. The
dashed lines indicate the transition temperatures.}
\end{figure}

In the case of the dynamic susceptibility as measured by ESR, the
contribution of the $e_g$ spin system can not be detected anymore,
when the hopping time becomes short compared to a Larmor period and
prevents the occurrence of the precessional motion of the
spin.\cite{Pake73} Therefore, the reduced effective moment observed
by ESR can be regarded as evidence that the time scale of the
electron hopping in YBaMn$_2$O$_6$ is much faster than the ESR
measuring frequency of 10$^{-10}$~sec. Hence, we are left with the
reduced $\chi_{\rm ESR}$ and an effective moment comparable to
5.48~$\mu_B$ of the Mn$^{4+}$ ions. We would like to point out that
structural investigations show that all Mn lattice site are
equivalent in the high-temperature state above $T_{\rm t}$,
supporting a picture of a system built up of Mn$^{4+}$ ions and a
highly mobile polaronic $e_g$-electron system, which accounts for
the metallic-like behavior.\cite{Williams05b}

Such a scenario certainly has to leave its footprints in the
relaxation dynamics of the spin system and, indeed, the quasi-linear
increase of the linewidth with temperature in the metallic regime
reminds of a Korringa-like relaxation mechanism with a slope of
about 2.125~Oe/K.\cite{Barnes81} In contrast to the classical
Korringa relaxation, where the localized spins are coupled to the
quasi-free conduction electrons, the relaxation in YBaMn$_2$O$_6$
must be of a more complicated nature due to the polaronic character
of the electrical conductivity. It has been argued that the
linewidth in such a case is also governed by the thermal activated
polaron hopping and follows the temperature dependence of the
conductivity. \cite{Shengelaya96,Shengelaya00} Although a thermally
activated behavior may appear as linear in a small temperature
range, the present data can only support a linear fit. Additional
measurements in a broader temperature range above $T_{\rm t}$ are
necessary to compare and check for different models of the spin
relaxation at high temperatures.

\subsection{Charge-order and charge fluctuations for $T<T_{\rm t}$}

As we leave the metallic-like regime for $T<T_{\rm t}$ we have to be
aware of the fact that $\chi_{\rm ESR}$ is still significantly lower
than $\chi_{\rm dc}$, but the difference is decreasing. This is an
indication that the hopping frequency of the $e_g$ electrons
approaches the ESR measurements frequency $\nu \approx 10^{10}$~Hz
and the $e_g$ spins start to participate in the resonance phenomena.
This slowing down of the $e_g$ electrons corresponds to a
localization process leading to a charge ordering into Mn$^{3+}$ and
Mn$^{4+}$ states. This growing contribution of Mn$^{3+}$ ions is
certainly reflected in the behavior of the effective $g$-factor and
the linewidth.

The $g$-factor is very sensitive to changes in the local symmetry of
the spin.\cite{Pake73,Barnes81,Zakharov06} Having six different Mn-O
bond lengths, the MnO$_6$ octahedra in \YBa~are heavily
distorted,\cite{Nakajima04} and one would expect the $g$-tensor to
be strongly anisotropic. Here, in the case of a polycrystalline
sample, the $g$-factor will be a statistical average of all
orientations. Hence, we will have to concentrate on the information
which can be obtained by the relative changes of the $g$ factors
upon temperature:

First, we want to refer to the properties of polycrystalline
La$_{1-x}$Sr$_x$MnO$_3$, where a similar change of the $g$-factor as
the one arising for $T<T_{\rm t}$ has been observed as a consequence
of a transition into a cooperatively Jahn-Teller distorted
phase.\cite{Tovar99,Ivanshin00} The reason for this shift is the
anisotropy arising from long-range orbital ordering of Mn$^{3+}$
ions as it was shown in single crystals of \LaSr. The main
parameters to describe the temperature dependence and anisotropy of
the $g$-factor were the local zero-field splitting (a measure of the
distortion of the MnO$_6$ octahedra) and the susceptibility of the
sample.\cite{Moreno01,Deisenhofer02} Clearly, in \YBa\; the
susceptibility changes dramatically and a considerable change of the
local Mn-O bond lengths with temperature has been reported,
too.\cite{Nakajima04,Williams05b}

\begin{figure}[t]
\centering
\includegraphics[width=\linewidth]{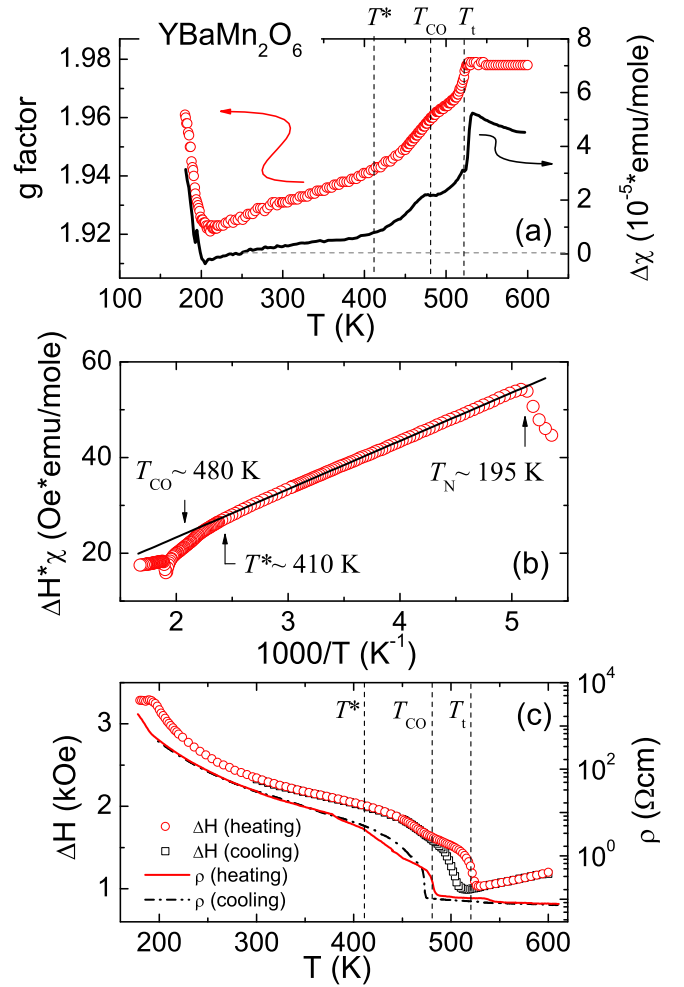}
\vspace{2mm} \caption[]{\label{dhanalysis} Analysis of the ESR
parameters in \YBa. (a): Temperature dependences of the $g$-factor
and of $\Delta \chi=\chi_{\rm dc}-\chi_{\rm ESR}$. (b): Temperature
dependence of $\Delta H \cdot \chi$ plotted vs. the inverse
temperature. (c): $\Delta H (T)$ compared to the temperature
dependence of the resistivity on a logarithmic scale.}
\end{figure}

A direct comparison of the temperature dependencies $\chi_{\rm ESR}$
and the $g$-factor in Fig.~\ref{YBMO} yields a linear behavior of
both in a broad range below $T_{\rm t}$, suggesting some correlation
of the two parameters. When approaching $T_{\rm CO}$ and $T_{\rm
t}$, however, the $g$-factor does not mimic the susceptibility
anymore. Assuming that the relative change in the $g$-factor towards
lower temperatures is due to the increasing number of Mn$^{3+}$, we
plot in Fig.~\ref{dhanalysis}(a) the $g$-factor together with
$\Delta \chi=\chi_{\rm dc}-\chi_{\rm ESR}$. The good agreement in
the whole temperature range up to $T_{\rm t}$ seems to justify our
assumptions up to now, $\Delta \chi$ and the $g$-factor are a
measure for the charge fluctuations and the charge ordering process
in our system.

It seems natural to expect the third ESR parameter, the linewidth
$\Delta H$, to reflect the localization process,
too.\cite{Heinrich04,Eremin05} In general, the linewidth is a
measure of the relaxation of the spin system. With a growing number
of Jahn-Teller active Mn$^{3+}$ for $T<T_{\rm t}$ the inhomogeneous
broadening due to the zero-field splitting by the crystal-field
becomes stronger and consequently the ESR line
broader.\cite{Deisenhofer03,Kochelaev03}

As pointed out by Huber {\it et al.},\cite{Huber99} the temperature
dependence of the linewidth in many
manganites\cite{Causa98,deisenhofer02,Atsarkin01} can be described
with the Kubo-Tomita approach\cite{Kubo54}
\begin{equation}
\Delta H(T) = \frac{\chi_0(T)}{\chi_{\rm dc}(T)}\Delta H_{\infty},
\label{Huber}
\end{equation}
as long as the systems are not too close to critical regions of
magnetic and structural phase transitions. Here, $\chi_0=C/T$
denotes the Curie susceptibility with the Curie constant $C=13N
g^2\mu^2_{\rm B} /4 k_{\rm B}=4.875\quad\mathrm{emu K/mol}$ of the
exchange-coupled Mn$^{3+}$/Mn$^{4+}$ system, and $\chi(T)$ is given
by the measured $dc$-susceptibility. Moreover, $\Delta H_{\infty}$
is a temperature independent parameter, which is usually identified
with the high-temperature limit of the ESR linewidth, if
$\chi_0(T)/\chi(T)\rightarrow 1$ for $T\rightarrow\infty$. Hence,
the temperature dependence of $\Delta H$ is dominated by $\chi_{\rm
dc}$ in this approximation. To check the validity of this formula we
plot $\Delta H(T)\cdot\chi(T)$ vs. reciprocal temperature in
Fig.~\ref{dhanalysis}(b), which results in a linear behavior for
200~K$<T< T^* \approx 410$~K. From a linear fit we derive $\Delta
H_{\infty}=2.1$~kOe which is in good agreement with the values
observed for other orbitally ordered phases in manganite systems,
where values of 2-3~kOe have been
found.\cite{Huber99,Tovar99,deisenhofer02} Astonishingly, the
resistivity\cite{Nakajima04} as a direct measure of the charge
ordering process may also be compared with the linewidth in
Fig.~\ref{dhanalysis}(c). Note that the resistivity is plotted on a
logarithmic scale. The similarity of the temperature dependence
encourages us to evoke the relation $\Delta H(T)\sim
\textrm{ln}(\rho (T))$ between the ESR linewidth and the resistivity
in the temperature range $T_{\rm N} < T < T_{\rm CO}$. To the best
of our knowledge, such a correlation between the ESR linewidth and
the resistivity has not been reported beforehand. Though
speculative, this observation points to a strong coupling of the
spin and charge degrees of freedom in this temperature range.

\section{Summary}

In summary, we found significant differences in the ESR properties
of \YBa\; between the high-temperature metallic-like regime for
$T>520$~K and the regime $T<520$~K. The ESR intensity in the
high-temperature regime was found to be only due to Mn$^{4+}$ ions
in contrast to the $dc$-susceptibility. This indicates that \YBa\;
represents a rare example of a system, where charge fluctuations
take place on a time scale shorter than the typical spin relaxation
and ESR measurement times. The linewidth for $T>520$~K seems to
follow a Korringa-like behavior. Towards lower temperatures the
system is dominated by the slowing down of these charge fluctuations
which influence the ESR absorption significantly. The ESR intensity
approaches the $dc$-susceptibility with decreasing temperature and
the two curves coincide only below a temperature $T^*\approx 410$~K,
which we interpret as the completion of the charge ordering process.
The temperature dependence of linewidth appears similar to the one
of the resistivity on a logarithmic scale.

We thank I. Leonov and D.L. Huber for stimulating discussions. This
work was supported by the German BMBF under Contract No. VDI/EKM
13N6917 and by the DFG within SFB 484 (Augsburg). The work of
D.~V.~Z. was supported by VW-Stiftung.

\end{document}